\documentclass[12pt]{iopart}

\usepackage{iopams}
\usepackage{graphicx}
\usepackage{amssymb}
\usepackage{dsfont}
\usepackage{bm}
\usepackage[square,sort&compress,numbers]{natbib}

\begin{document}

\title{Directional correlations in quantum walks with two particles}

\author{M. \v Stefa\v n\'ak$^{(1)}$, S. M. Barnett$^{(2)}$, B. Koll\'ar$^{(3)}$, T. Kiss$^{(3)}$ and I. Jex$^{(1)}$}

\address{$^{(1)}$ Department of Physics, Faculty of Nuclear Sciences and Physical Engineering, Czech Technical University in Prague, B\v
rehov\'a 7, 115 19 Praha 1 - Star\'e M\v{e}sto, Czech Republic}
\address{$^{(2)}$ Department of Physics, University of Strathclyde, 107 Rottenrow, Glasgow, G4 0NG, Scotland, U.K.}
\address{$^{(3)}$ Department of Quantum Optics and Quantum Information, Research Institute for Solid State Physics and Optics, Hungarian Academy of Sciences, Konkoly-Thege u. 29-33, H-1121 Budapest, Hungary}


\date{\today}

\begin{abstract}
Quantum walks on the line with a single particle possess a classical analog. Involving more walkers opens up the possibility to study collective quantum effects, such as many particle correlations. In this context, entangled initial states and indistinguishability of the particles play a role. We consider directional correlations between two particles performing a quantum walk on a line. For non-interacting particles we find analytic asymptotic expressions and give the limits of directional correlations. We show that introducing $\delta$-interaction between the particles, one can exceed the limits for non-interacting particles.
\end{abstract}

\maketitle


\section{Introduction}
\label{sec:1}

Quantum walks were introduced \cite{aharonov} as a generalization of a classical random walk \cite{hughes} to a unitary evolution of a quantum particle. The time evolution can be either discrete \cite{meyer} or continuous \cite{farhi}. The connection between discrete-time and continuous-time quantum walks has been established for a walk on a line \cite{strauch,chandra:08} and, more recently, for walks on arbitrary graphs \cite{childs:walks}. It has been shown that both continuous \cite{childs:09} and discrete time \cite{lovett:09} quantum walks can be regarded as a universal computational primitive. Continuous time quantum walks have been extensively studied in the context of coherent energy transfer in networks \cite{muelken:prl}. Both continuous and discrete time quantum walks have found a promising application in designing quantum algorithms \cite{kempe}. Indeed, a number of algorithms based on quantum walks have been proposed \cite{shenvi:2003,ambainis:2003,childs:04,kendon:2006,aurel:2007,magniez,Reitzner,vasek,Tanner:search:2009,Tanner:search:2010}, for a review see \cite{santha:2008}.

Various properties of quantum walks have been analyzed, in particular their asymptotic behavior \cite{Grimmett,konno:2002b,konno:2005b} and the effect of the initial conditions \cite{tregenna,chandrashekar:2007,miyazaki}, for a review see \cite{konno:book}. Due to the wave-nature of quantum walks a number of counter-intuitive phenomena has been observed, including infinite hitting times \cite{krovi:2006a,krovi:2006b} and localization \cite{mackay,konno:loc:2004,konno:loc:2005,konno:loc:2005b,konno:loc:2008,konno:loc:2010,shikano:loc}. The properties of random walks on infinite regular lattices are closely related to the dimensionality of the lattice. It is well known that a classical random walk returns to the origin with certainty in dimension one and two while in higher dimensions the probability of return (P\'olya number) is strictly less than unity \cite{polya}. For the discrete-time quantum walk the recurrence properties are determined not only by the dimension but also by the initial state and the coin operator leading to rich behavior \cite{stef:prl,stef:pra,stef:njp,kiss:recurrence,kollar,chandra:rec}. The closely related property of persistence has been studied in \cite{Goswani:pers}.

The extensive theoretical studies have stimulated the search for experimental implementations of quantum walks. Various schemes based on ion traps \cite{travaglione:walk:ion}, optical lattices \cite{dur:walk:lattice,eckert:walk:optraps}, cavity quantum electrodynamics \cite{sanders:walk:cqed}, optical cavities \cite{knight:walk:optcav} or Bose-Einstein condensate \cite{chandra:walk:bec} have been proposed. Recently, discrete time quantum walk on the line has been realized in a variety of physical systems including cold atoms \cite{karski}, trapped ions \cite{schmitz,Zahringer} and photons \cite{Schreiber,Broome}.

Most of the studies to date considered quantum walks with a single particle. A natural extension of the field of quantum walks is to involve more particles. This unlocks the additional features offered by quantum mechanics such as entanglement and indistinguishability which are not available in classical random walks. Quantum walk on a line with two entangled particles has been introduced in \cite{omar} and the meeting problem in this model has been analyzed \cite{stef:meeting}. A physical implementation of this model based on linear optics has been proposed \cite{pathak}. Quantum walks with two particles have been applied to the graph isomorphism problem \cite{shiau,gamble}. Entanglement generation in a special two-particle quantum walk on a line has been investigated in \cite{alles}. Recently, the first successful experiment with two particles on a line has been reported \cite{OBrien}. A framework for multi-particle quantum walks on rather arbitrary graphs has been proposed in \cite{rohde}. The study of quantum walks with more particles on the line is motivated by the fact that the single-particle walk in this case can be considered as a classical interference phenomenon \cite{knight}. We note that walks on higher-dimensional lattices cannot be considered classical
in this sense, since the resources needed to simulate the quantum walk scale exponentially.

In this paper, we investigate the non-classical effects in the two-particle discrete-time quantum walk on the line by asking the question: How is the directional correlation affected by the quantum nature of the particles? In particular, we analyze the probability $P_s$ of finding both particles on the same (negative or positive) half-line. We derive analytical expressions for the asymptotic value of this probability in dependence on the initial coin state. Classically, a symmetric random walk has a fixed value of the probability $P_s$ equal to 1/2. We first consider two quantum particles on a line starting the walk in a separable state. We determine the limits for the directional correlations and show that, for any value within these limits, one can design a corresponding separable initial state. Next, we prove that the bounds cannot be exceeded by considering entanglement in the initial state. On the other hand, introducing quantum walks with $\delta$-interactions, we show that the directional correlations can be increased above the limit for non-interacting particles.

Our paper is organized as follows: we briefly review the quantum walk on a line with one and two non-interacting particles in Section~\ref{sec:2} and introduce the probability to be on the same side of the lattice $P_s$. In Section~\ref{sec:3} we analyze the probability $P_s$ for separable initial states. Entangled initial states are considered in Section~\ref{sec:4}. In Section~\ref{sec:5} we study the influence of the indistinguishability on the probability
$P_s$. In Section~\ref{sec:6} we introduce the concept of $\delta$-interacting quantum walks to break the limits of non-interacting quantum walks. We summarize our results in Section~\ref{sec:7}.

\section{Quantum walk on a line with one and two particles}
\label{sec:2}

Let us first briefly review the quantum walk of a single particle on a line (see e.g.  Ref. \cite{kempe:qw:overview} for a more detailed introduction). The Hilbert space of the particle is given by a tensor product
$$
\mathcal{H} = \mathcal{H}_P\otimes\mathcal{H}_C
$$
of the position space
$$
{\cal H}_P = \ell^2(\mathds{Z}) = \textrm{Span}\left\{|m\rangle|\ m\in\mathds{Z}\right\}
$$
and the two-dimensional coin space
$$
\mathcal{H}_C = \textrm{Span}\left\{|L\rangle,|R\rangle\right\}\,.
$$
We consider a particle starting the quantum walk from the origin, i.e. the initial state has the form
$$
|\psi(0)\rangle = |0\rangle\otimes|\psi_C\rangle,
$$
where $|\psi_C\rangle$ denotes the initial state of the coin. After $t$ steps of the quantum walk the state of the particle is given by
\begin{equation}
|\psi(t)\rangle \equiv \sum_m \left(\frac{}{}\psi_L(m,t)|m\rangle|L\rangle + \psi_R(m,t)|m\rangle|R\rangle\right) = U^t|\psi(0)\rangle\,,
\label{qw:time:evol}
\end{equation}
where the unitary propagator $U$ has the form
\begin{equation}
\label{1part:prop}
U = S \left(I \otimes C\right) \,.
\end{equation}
The coin operator $C$ flips the state of the coin before the particle is displaced. In principle, $C$ can be an arbitrary unitary operation on the coin space $\mathcal{H}_C$. We choose the most studied case of the Hadamard coin, denoted by $C_H$, which is defined by its action on the basis states
$$
C_H|L\rangle = \frac{1}{\sqrt{2}}(|L\rangle + |R\rangle),\quad
C_H|R\rangle = \frac{1}{\sqrt{2}}(|L\rangle - |R\rangle).
$$
After the coin flip the step operator $S$ displaces the particle from its current position according to its coin state
$$
S|m\rangle|L\rangle \longrightarrow |m-1\rangle|L\rangle,\quad S|m\rangle|R\rangle \longrightarrow |m+1\rangle|R\rangle.
$$
The coefficients $\psi_{L,(R)}(m,t)$ in (\ref{qw:time:evol}) represent the probability amplitudes of finding the particle at position $m$ after $t$ steps of the quantum walk with the coin state $|L(R)\rangle$. The probability distribution generated by the quantum walk is given by
$$
p(m,t) = |\langle m|\langle L|\psi(t)\rangle|^2 + |\langle m|\langle R|\psi(t)\rangle|^2 = |\psi_L(m,t)|^2 + |\psi_R(m,t)|^2\,.
$$

The extension of the formalism described above to two distinguishable particles has been given in \cite{omar}. One should consider the bipartite Hilbert state as a tensor product
$$
\mathcal{H}_{1 2} = \mathcal{H}_{1}\otimes\mathcal{H}_{2}
$$
of the single particle Hilbert spaces. We consider non-interacting particles, i.e. their time evolution is independent. Hence, the propagator of the two-particle quantum walk can be written in a factorized form
\begin{equation}
\label{qw:2part:factor}
U_{1 2} = U_{1} \otimes U_{2},
\end{equation}
where $U_1$ $(U_2)$ is the propagator of the first (second) particle given by Eq. (\ref{1part:prop}). Note that this factorized time evolution cannot increase entanglement between the particles. In this paper we consider particles starting from the same lattice point (the origin). Hence, the initial state of the two-particle quantum walk has the shape
$$
|\Psi(0)\rangle = |0,0\rangle \otimes |\Psi_C\rangle,
$$
where $|\Psi_C\rangle$ is the initial coin state of the two particles.

Let us first consider the case when the initial coin state is separable, i.e.
\begin{equation}
\label{psi:c:fact}
|\Psi_C\rangle = |\psi_1\rangle\otimes|\psi_2\rangle\,.
\end{equation}
Since entanglement is not induced in the process of time evolution, the two-particle state remains factorized and the joint probability distribution $p(m,n,t)$ of finding the first particle at the $m$th and the second at the $n$th site at time $t$ is reduced to the product of single particle distributions
\begin{equation}
\label{2part:dist:fact}
p(m,n,t) = p_1(m,t)\cdot p_2(n,t)\, .
\end{equation}
Here, $p_i(m,t)$ is the probability distribution of a single-particle quantum walk given that the initial coin state was $|\psi_i\rangle$. Hence, the two-particle quantum walk with initially separable coin state is fully determined by the single-particle quantum walk.

We turn to the situation when the initial coin state $|\Psi_C\rangle$ does not factorize. In such a case, the joint probability distribution $p(m,n,t)$ cannot be written in a product form (\ref{2part:dist:fact}). Nevertheless, we can map the two-particle walk on a line to a quantum walk of a single particle on a square lattice. Indeed, we can write the two-particle propagator (\ref{qw:2part:factor}) in the following form
\begin{equation}
\label{2part:U}
U_{1 2} = S_{1 2} (I_{P_{12}} \otimes (C_H \otimes C_H)),
\end{equation}
where $I_{P_{12}}$ is the identity on the joint position space and the joint step operator $S_{1 2}$ is given by the tensor product of the single particle step operators $S_i$. The relation (\ref{2part:U}) implies that we can consider the two-particle propagator $U_{1 2}$ as a propagator of single-particle walk on a plane with the coin given by the tensor product of two Hadamard operators. Hence, the two quantum walks in consideration are equivalent. This correspondence allows us to treat the joint probability distribution of the two-particle walk with the tools developed for the single-particle quantum walks.

Finally, let us briefly comment on a quantum walk with indistinguishable particles. It is natural to use the second quantization formalism. We denote the bosonic creation operators by $\hat{a}_{(m,i)}^\dagger$ and the fermionic creation operators by $\hat{b}_{(n,j)}^\dagger$, e.g. $\hat{a}_{(m,i)}^\dagger$ creates one bosonic particle at position $m$ with the internal state $|i\rangle$, $i = L,R$. The dynamics of the quantum walk with indistinguishable particles is defined on a one-particle level, i.e. a single step is given by the following transformation of the creation operators
$$
\hat{a}_{(m,L)}^\dagger \longrightarrow \frac{1}{\sqrt{2}}\left(\hat{a}_{(m-1,L)}^\dagger + \hat{a}_{(m+1,R)}^\dagger\right)\ ,\quad  \hat{a}_{(m,R)}^\dagger \longrightarrow \frac{1}{\sqrt{2}}\left(\hat{a}_{(m-1,L)}^\dagger - \hat{a}_{(m+1,R)}^\dagger\right)\,,
$$
for bosonic particles, similarly for fermions. The difference is that the bosonic operators fulfill the commutation relations
\begin{equation}
\label{commut}
\left[\hat{a}_{(m,i)},\hat{a}_{(n,j)}\frac{}{}\right] = 0\,,\qquad \left[\hat{a}_{(m,i)},{\hat{a}_{(n,j)}}^\dagger\right] = \delta_{mn}\delta_{ij}\,,
\end{equation}
while the fermionic operators satisfy the anti-commutation relations
\begin{equation}
\label{anticommut}
\left\{\hat{b}_{(m,i)},\hat{b}_{(n,j)}\frac{}{}\right\} = 0\,,\qquad \left\{\hat{b}_{(m,i)},\hat{b}^\dagger_{(n,j)}\right\} = \delta_{mn}\delta_{ij}\,.
\end{equation}
Since the dynamics is defined on a single-particle level, one can describe the state of the two indistinguishable particles after $t$ steps of the quantum walk in terms of the single-particle probability amplitudes (see Ref. \cite{stef:meeting} for a more detailed discussion).

In the present paper we focus on the directional correlations between the two particles. We quantify this property by the probability $P_s$ that both particles are found after $t$ steps of the quantum walk on the same side of the line. For distinguishable particles it is given by
\begin{equation}
\label{P:same}
P_s(t) = \sum_{m=-t}^0\sum_{n=-t}^0 p(m,n,t) + \sum_{m=1}^t\sum_{n=1}^t p(m,n,t)\,.
\end{equation}
For indistinguishable particles $p(m,n,t)\equiv p(n,m,t)$, i.e. these two probabilities correspond to the same physical event. Hence, the sums in (\ref{P:same}) have to be restricted over an ordered pair $(m,n)$ with $m\geq n$, i.e.
\begin{equation}
\label{P:same:indist}
P_s(t) = \sum_{n=-t}^0\left(\sum_{m=n}^0 p(m,n,t)\right) + \sum_{n=1}^t\left(\sum_{m=n}^t p(m,n,t)\right)\,.
\end{equation}
In particular, we will be interested in the asymptotic limits of the probability $P_s$ in its dependence on the initial coin state of the two particles. We consider both separable and entangled coin states, as well as indistinguishability of the particles, in the following Sections.


\section{Separable initial states}
\label{sec:3}

Let us now specify the probability $P_s$ for two distinguishable particles which start the quantum walk with a separable coin state (\ref{psi:c:fact}). As discussed in the previous Section the joint probability distribution $p(m,n,t)$ factorizes (\ref{2part:dist:fact}). Therefore, the probability to be on the same side of the lattice $P_s$ simplifies into
\begin{equation}
\label{P:factor}
P_s(t) = P_1^-(t)\cdot P^-_2(t) + P^+_1(t)\cdot P^+_2(t)\,,
\end{equation}
Here we have denoted by $P^\pm_i(t)$ the probability that the particle which have started the quantum walk with the coin state $|\psi_i\rangle$ is on the positive or negative half-axis after $t$ steps, i.e.
$$
P^-_i(t) = \sum_{m=-t}^0 p_i(m,t),\quad P^+_i(t) = \sum_{m=1}^t p_i(m,t).
$$

\begin{figure}
\begin{center}
\includegraphics[width=0.7\textwidth]{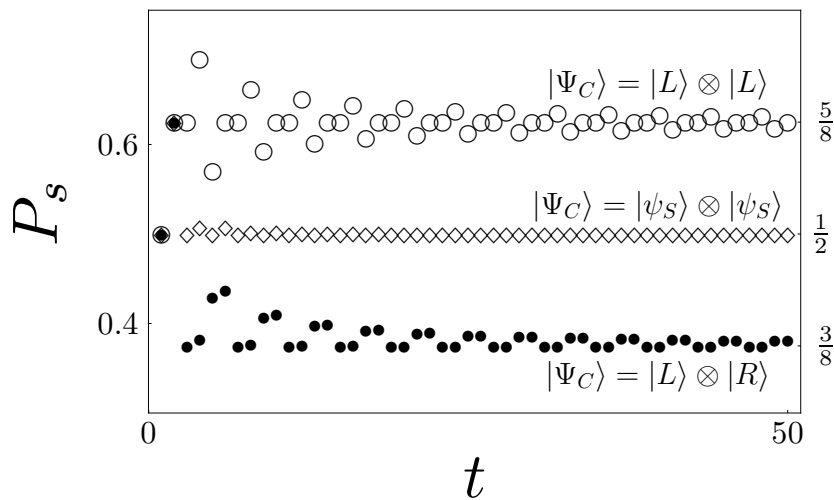}
\caption{The probability $P_s$ that two distinguishable particles performing a quantum walk on a line end on the same side as a function of time. Both particles start the quantum walk from the origin. As the initial coin state $|\Psi_C\rangle$ we choose one of the three factorized states - $(\it i)$ $|L\rangle$ for the first particle and $|R\rangle$ for the second particle (black dots), $(\it ii)$ $|L\rangle$ for both particles (open circles), and $(\it iii)$ $|\psi_S\rangle \equiv \frac{1}{\sqrt{2}}\left(|L\rangle+i|R\rangle\right)$ for both particles (open diamonds). We find that for the initial coin state $(\it i)$ the particles are more likely to be on the opposite side, since $P_s<1/2$. Indeed, due to the choice of the coin state $|LR\rangle$ the probability distribution of the first particle is biased to the left while the probability distribution of the second particle is biased to the right. On the other hand, for the initial state $|LL\rangle$ both probability distributions are biased to the left. Hence, the particles are more likely to be found on the same side. Finally, for the initial state $(\it iii)$ which results in the symmetric single-particle probability distribution the particles are equally likely to be on the same or the opposite side of the line. The asymptotic values of $P_s$ for all three initial states are in agreement with the analytic estimation of Eq.(\ref{P:s:factor:special}).}
\label{fig1}
\end{center}
\end{figure}

In Figure~\ref{fig1} we plot the course of the probability $P_s(t)$ with the number of steps $t$. To unravel the dependence on the initial coin state $|\Psi_C\rangle$ we consider three cases - $(\it i)$ $|\Psi_C\rangle=|L\rangle\otimes|R\rangle$ (black dots), $(\it ii)$ $|\Psi_C\rangle=|L\rangle\otimes|L\rangle$ (open circles), and $(\it iii)$ $|\Psi_C\rangle=\frac{1}{\sqrt{2}}\left(|L\rangle+i|R\rangle\right)\otimes\frac{1}{\sqrt{2}}\left(|L\rangle+i|R\rangle\right)$ (open diamonds). We find that after some initial oscillations the probability $P_s$ quickly approach steady values which are determined by the initial coin state.


Let us now determine the asymptotic value of the probability $P_s$ in dependence of the initial coin state. Consider a general separable coin state of the form
$$
|\Psi_C\rangle = \left(a_1|L\rangle + b_1|R\rangle\right)\otimes\left(a_2|L\rangle + b_2|R\rangle\right)\,.
$$
The asymptotic probability distribution for a single particle is given by \cite{konno:2002b}
\begin{equation}
\label{approx:factor}
p(x,t,a_i,b_i)=\frac{1-\frac{x}{t}((a_i+b_i) \overline{a_i}+(a_i-b_i)\overline{b_i})}{\pi t\sqrt{1-2\frac{x^2}{t^2}}(1-\frac{x^2}{t^2})}\,.
\end{equation}
The probability that the particle is on the negative or positive half-axis is obtained by integrating the probability density over the corresponding interval
\begin{eqnarray}
\label{P:pm:factor}
\nonumber P^-_i(a_i,b_i) & = & \int\limits_{-\frac{t}{\sqrt{2}}}^0 p(x,t,a_i,b_i) dx = \frac{1}{4}(2+((a_i+b_i)\overline{a_i}+(a_i-b_i)\overline{b_i}))\,,\\
P^+_i(a_i,b_i) & = & \int\limits_{0}^{\frac{t}{\sqrt{2}}} p(x,t,a_i,b_i) dx = \frac{1}{4}(2-((a_i+b_i)\overline{a_i}+(a_i-b_i)\overline{b_i}))\,.
\end{eqnarray}
Note that within the approximation of Eq. (\ref{approx:factor}) the resulting integrals are time-independent, i.e. we immediately obtain the asymptotic values of the probabilities $P^\pm_i$. This is due to the fact that the asymptotic probability density depends only on the ratio $x/t$.

Inserting the results of (\ref{P:pm:factor}) into the Eq.(\ref{P:factor}) we find that the probability $P_s$ is given by
\begin{eqnarray}
\label{P:s:factor}
\nonumber P^{(sep)}_s = \frac{1}{8}\left(\frac{}{}4+\left((a_1+b_1)\overline{a_1}+(a_1-b_1)\overline{b_1}\right)\left((a_2+b_2) \overline{a_2}+(a_2-b_2)\overline{b_2}\right)\frac{}{}\right)\,.\\
\end{eqnarray}
In particular, for the initial states $(\it i-iii)$ considered in Figure~\ref{fig1}, we find the asymptotic values
\begin{eqnarray}
\label{P:s:factor:special}
\nonumber P^{(LR)}_s & \equiv & P_s(1,0,0,1) = \frac{3}{8}\,, \quad P^{(LL)}_s \equiv P_s(1,0,1,0) = \frac{5}{8}\,,\\
P^{(S)}_s & \equiv & P_s(\frac{1}{\sqrt{2}},\frac{i}{\sqrt{2}},\frac{1}{\sqrt{2}},\frac{i}{\sqrt{2}}) = \frac{1}{2}\,.
\end{eqnarray}
These results are in perfect agreement with the numerical simulations presented in Figure~\ref{fig1}.

Let us now analyze the probability $P^{(sep)}_s$ in more detail. First, we recast the formula (\ref{P:s:factor}) in a simpler form by a change of the basis of the coin space. Consider the basis formed by the eigenstates of the Hadamard coin
$$
C_H |\chi^\pm\rangle = \pm|\chi^\pm\rangle\,,
$$
which have the following expression in the standard basis
\begin{equation}
\label{basis:had}
|\chi^\pm\rangle = \frac{\sqrt{2\pm\sqrt{2}}}{2}|L\rangle \pm \frac{\sqrt{2\mp\sqrt{2}}}{2}|R\rangle\,.
\end{equation}
We decompose the single-particle coin state in the Hadamard basis
$$
|\psi_i\rangle = h_i^+|\chi^+\rangle + h_i^-|\chi^-\rangle\,.
$$
From the expression (\ref{basis:had}) we find the transformation between the coefficients in the standard and the Hadamard basis
$$
a_i = \frac{\sqrt{2+\sqrt{2}}}{2} h_i^+ +\frac{\sqrt{2-\sqrt{2}}}{2} h_i^-\,,\quad
b_i = \frac{\sqrt{2-\sqrt{2}}}{2} h_i^+ -\frac{\sqrt{2+\sqrt{2}}}{2}  h_i^-\,.
$$
With the help of these relations we find that the formula (\ref{P:s:factor}) for the probability $P^{(sep)}_s$ simplifies in the Hadamard basis into
\begin{equation}
\label{Ps:sep:had}
P^{(sep)}_s = \frac{1}{4}\left(2+(2\left|h_1^+\right|^2-1) (2\left|h_2^+\right|^2-1)\right)\,.
\end{equation}
Here we have used the normalization of the single-particle coin state $|\psi_i\rangle$, i.e. the condition
\begin{equation}
\label{norm:fact}
|h_i^+|^2 + |h_i^-|^2 = 1\,.
\end{equation}

\begin{figure}
\begin{center}
\includegraphics[width=0.7\textwidth]{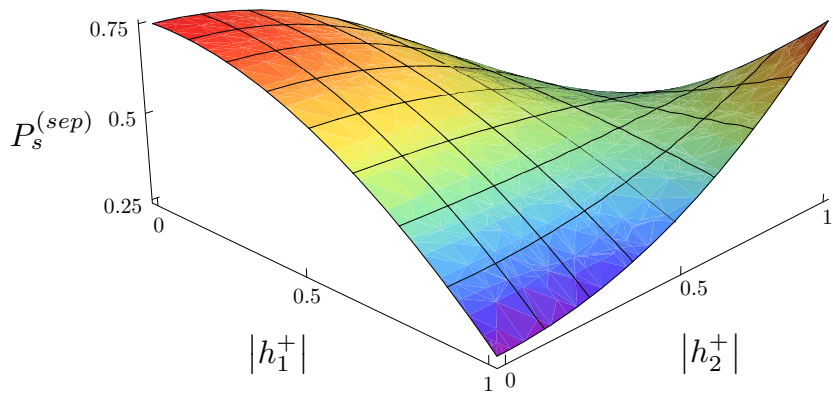}
\caption{The probability to be on the same side $P^{(sep)}_s$ in its dependence on the coefficients of the initial coin states. The parameters $h_i^+$ are given by the overlap of the coin state $|\psi_i\rangle$ with the eigenstate $|\chi^+\rangle$ of the Hadamard coin. We find that the probability $P^{(sep)}_s$ reaches the maximum value 3/4 when both $h_{1,2}^+$ equals either zero or one. The minimum value 1/4 is obtained if one the $h_i^+$ is zero while the other one is unity.}
\label{fig1:2}
\end{center}
\end{figure}

\begin{figure}
\begin{center}
\includegraphics[width=0.6\textwidth]{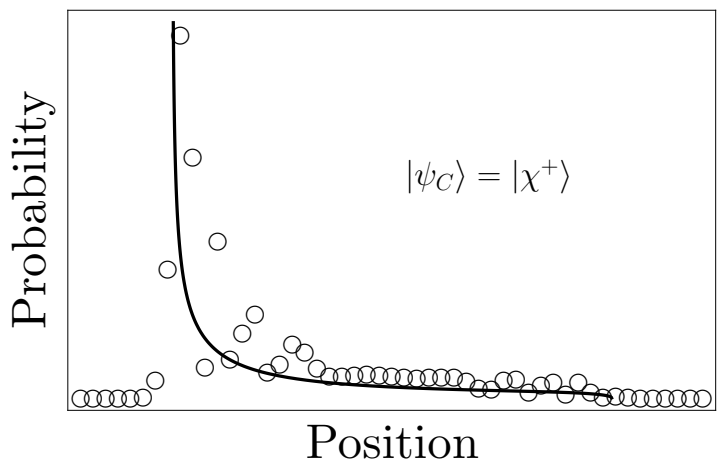}
\caption{Single-particle probability distribution for the initial coin state $|\psi_C\rangle = |\chi^+\rangle$. We find only one peak on the left side of the lattice, the peak on the right side has disappeared. Consequently, the resulting probability distribution is maximally biased towards left. Choosing the initial coin state as $|\psi_C\rangle = |\chi^-\rangle$ will flip the plot around the origin and the resulting probability distribution will be maximally biased to the right.}
\label{fig1:3}
\end{center}
\end{figure}

We display the probability to be on the same side $P^{(sep)}_s$ in its dependence on the parameters $h_i^+$ in Figure~\ref{fig1:2}.
We find that $P^{(sep)}_s$ reaches the maximum value 3/4 provided that both $h_i^+$ equals zero or unity, i.e. when both particles start the walk in the same eigenstate of the Hadamard coin. Indeed, starting the single-particle walk in the eigenstate $|\chi^+\rangle$ ($|\chi^-\rangle$) leads to a probability distribution which is maximally biased towards left (right). We illustrate this feature in Figure~\ref{fig1:3}. Note that this effect has been identified numerically in \cite{tregenna}. Hence, when both particles start the walk in the same eigenstate of the Hadamard coin, their probability distributions are maximally biased towards the same direction and, consequently, the particles are the most likely to be on the same side. On the other hand, if the particles start the walk in the different eigenstates (e.g. the first one in $|\chi^+\rangle$ and the second one in $|\chi^-\rangle$, which corresponds to $h_1^+=1$ and $h_2^+=0$), the probability distributions are maximally biased in the opposite directions. In such a case, the particles are the most likely to be on the opposite side of the lattice and $P^{(sep)}_s$ reaches the minimum 1/4.


\section{Entangled initial states}
\label{sec:4}

Let us now analyze the probability that the particles will be on the same side of the lattice $P_s$ for the initial coin states $|\Psi_C\rangle$ which are not factorized. We follow two approaches. First, we analyze the particular case of maximally entangled Bell states. We express the two-particle state in terms of single-particle amplitudes. In this way, we decompose the joint probability distribution into single-particle distributions plus an interference term. We then use the results of the previous section to find the asymptotic value of the probability $P_s$. By this approach we emphasize the role of the interference of probability amplitudes. Second, we employ the equivalence between the two-particle walk on a line and single-particle walk on a square lattice discussed in Section~\ref{sec:2}. This correspondence allows us to use the tools developed for the quantum walks with a single particle, namely the weak limit theorems \cite{Grimmett}, to find the asymptotic probability density for the two-particle walk on a line. We leave the details of the calculation for the \ref{appendix}. With the explicit form of the probability density we finally derive the asymptotic value of the probability $P_s$ for an arbitrary two-particle coin state.

We start by examining the particular case of maximally entangled Bell states
\begin{equation}
\label{bell}
|\psi^\pm\rangle = \frac{1}{\sqrt{2}}\left(|LR\rangle \pm |RL\rangle\right)\,,\quad |\phi^\pm\rangle = \frac{1}{\sqrt{2}}\left(|LL\rangle \pm |RR\rangle\right)\,.
\end{equation}
Obviously, the joint probability distribution $p(m,n,t)$ is no longer a product of the single-particle probability distributions. However, we can still express it in terms of the single-particle probability amplitudes. Let us denote by $\psi^{(L)}_i(m,t)$ the amplitude of the particle being after $t$ steps at the position $m$ with the coin state $|i\rangle$, $i = L,R$, provided that the initial coin state was $|L\rangle$. Similarly, let $\psi^{(R)}_i(m,t)$ be the amplitude for the initial coin state $|R\rangle$. With this notation we express the joint probability distributions generated by quantum walk of two particles with initially entangled coins by
\begin{eqnarray}
\nonumber p^{(\psi^\pm)}(m,n,t) & = & \frac{1}{2} \sum_{i,j=L,R}\left|\psi_i^{(L)}(m,t)\psi_j^{(R)}(n,t)\pm \psi_i^{(R)}(m,t)\psi_j^{(L)}(n,t)\right|^2\,,\\
\nonumber p^{(\phi^\pm)}(m,n,t) & = & \frac{1}{2} \sum_{i,j=L,R}\left|\psi_i^{(L)}(m,t)\psi_j^{(L)}(n,t)\pm\psi_i^{(R)}(m,t)\psi_j^{(R)}(n,t)\right|^2\,,
\end{eqnarray}
where the superscript indicates the initial coin state. We now make use of the fact that the amplitudes $\psi_i^{(L,R)}(m,t)$ are real valued. Indeed, both the Hadamard coin and the initial states have only real entries. Hence, the amplitudes cannot attain any imaginary part during the time evolution. Therefore, we can replace the absolute values by simple brackets and expand the joint probability distributions in the form
\begin{eqnarray}
\nonumber p^{(\psi^\pm)}(m,n,t)  & =  & \frac{1}{2}\left(p^{(L)}(m,t)p^{(R)}(n,t) + p^{(R)}(m,t)p^{(L)}(n,t)\frac{}{}\right) \pm\\
\nonumber & & \pm \varphi(m,t)\varphi(n,t)\,,\\
\nonumber p^{(\phi^\pm)}(m,n,t)  & =  & \frac{1}{2}\left(p^{(L)}(m,t)p^{(L)}(n,t) + p^{(R)}(m,t)p^{(R)}(n,t)\frac{}{}\right) \pm\\
& & \pm \varphi(m,t)\varphi(n,t)\,.
\label{Pent}
\end{eqnarray}
Here, we have used the notation
$$
\varphi(m,t) = \psi_L^{(L)}(m,t)\psi_L^{(R)}(m,t) + \psi_R^{(L)}(m,t)\psi_R^{(R)}(m,t)
$$
to shorten the formulas. When we insert the expressions (\ref{Pent}) into the definition (\ref{P:same}) of the probability $P_s$ we find that the later one can be written in the form
$$
P^{(\psi^\pm)}_s(t) = P^{(LR)}_s(t) \pm I(t),\ P^{(\phi^\pm)}_s(t) = P^{(LL)}_s(t) \pm I(t)\,.
$$
The interference term $I(t)$ is given by
$$
I(t) = \left(\varphi^-(t)\frac{}{}\right)^2 + \left(\varphi^+(t)\frac{}{}\right)^2\,,
$$
where we have denoted
$$
\varphi^-(t) = \sum_{m=-t}^0\varphi(m,t),\quad \varphi^+(t) = \sum_{m=1}^t\varphi(m,t)\,.
$$

Let us now turn to the asymptotic values of $P_s$ in dependence on the choice of the Bell state. The limits of $P^{(LR)}_s$ and $P^{(LL)}_s$ are given in (\ref{P:s:factor:special}). We obtain the asymptotic value of the interference term $I(t)$ from the numerical simulation, which indicates
$$
I(t\rightarrow+\infty) = \frac{1}{8}\,.
$$
Finally, for the limiting values of the probability $P_s$ we find
\begin{equation}
P_s^{(\psi^+)} = \frac{1}{2},\quad P_s^{(\psi^-)} = \frac{1}{4},\quad P_s^{(\phi^+)} = \frac{3}{4},\quad P_s^{(\phi^-)} = \frac{1}{2}\,.
\label{P:s:o:ent}
\end{equation}

\begin{figure}
\begin{center}
\includegraphics[width=0.7\textwidth]{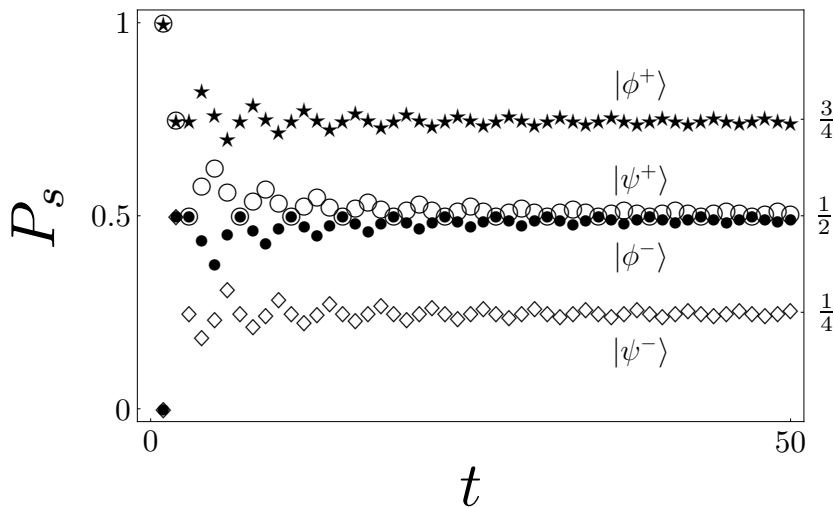}
\caption{The probability that two distinguishable particles performing a quantum walk on a line with initially entangled coins end on the same side as a function of time. Both particles start the quantum walk from the origin. As the initial coin state $|\Psi_C\rangle$ we choose one of the Bell states (\ref{bell}). For the Bell states $|\psi^+\rangle$ (open circles) and $|\phi^-\rangle$ (black dots) the particles are equally likely to be found on the same or on the opposite side of the line in the long time limit. For finite times they are more likely to be on the same side for $|\psi^+\rangle$ and more likely on the opposite side for $|\phi^-\rangle$. For the other two Bell states $|\psi^-\rangle$ (open diamonds) and $|\phi^+\rangle$ (stars) the differences remain in the asymptotic limit. The particles are more likely to be on the opposite side for the singlet state $|\psi^-\rangle$ and more likely to be on the same side for $|\phi^+\rangle$. The asymptotic values of the probability $P_s$ agree with the findings of (\ref{P:s:o:ent}).}
\label{fig2}
\end{center}
\end{figure}

We display the dependence of $P_s$ on the number of steps and the choice of the initial coin state in Figure~\ref{fig2}. We find that the probability $P_s$ quickly approach the steady values, similarly as for the factorized coin states which we have shown in Figure~\ref{fig1}. For $|\psi^+\rangle$ (open circles) and $|\phi^-\rangle$ (black dots) the particles are asymptotically equally likely to be on the same or on the opposite side. For the Bell state $|\phi^+\rangle$ (stars) the particles are more likely to be on the same side of the line. Finally, for the singlet state $|\psi^-\rangle$ (open diamonds) the particles are more likely to be on the opposite side. The asymptotic values of the probabilities $P_s$ are in agreement with the results of (\ref{P:s:o:ent}).


After we have analyzed the particular case of the Bell states we turn to a general initial coin state. As in the previous Section, we make use of the asymptotic probability density $p(x_1,x_2,t)$ and replace the sums in (\ref{P:same}) by integrals. We derive the explicit form of the asymptotic probability density in the \ref{appendix}. Performing the integrations we arrive at the following expression
$$
P_s = \frac{1}{4}\left(\frac{}{} 2 + |h_{(++)}|^2+|h_{(--)}|^2-|h_{(+-)}|^2-|h_{(-+)}|^2\right)
$$
for the probability to be on the same side. Here we have denoted by $h_{(\alpha\beta)}$ the coefficients of the decomposition of the initial coin state $|\Psi_C\rangle$ into the basis formed by the tensor product of the eigenvectors $|\chi^\pm\rangle$ of the Hadamard coin $C_H$, i.e.
\begin{equation}
|\Psi_C\rangle = \sum_{\alpha,\beta = \pm} h_{(\alpha\beta)}|\chi^\alpha\rangle|\chi^\beta\rangle.
\label{init:2part:had}
\end{equation}
Finally, using the normalization condition for the initial coin state $|\Psi_C\rangle$
$$
|h_{(++)}|^2 + |h_{(-+)}|^2 + |h_{(+-)}|^2 + |h_{(--)}|^2 = 1,
$$
we can simplify the expression for the probability $P^{(ent)}_s$ into the form
\begin{equation}
\label{Ps:ent}
P^{(ent)}_s = \frac{1}{4}\left(\frac{}{} 1 + 2(|h_{(++)}|^2 + |h_{(--)}|^2) \right)\,.
\end{equation}

\begin{figure}
\begin{center}
\includegraphics[width=0.7\textwidth]{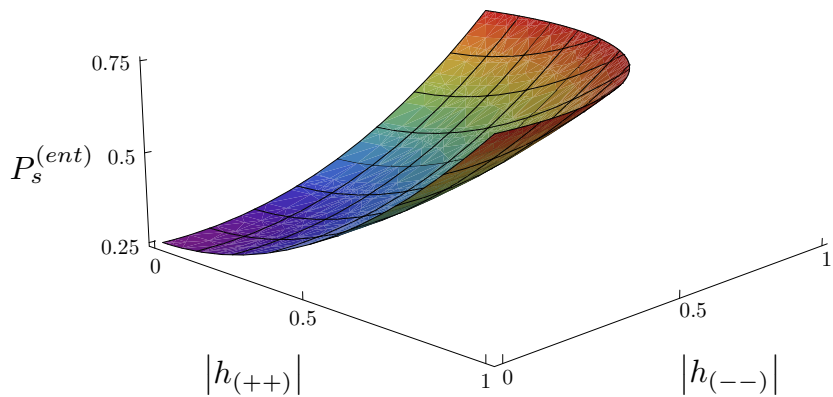}
\caption{The probability to be on the same side of the lattice $P^{(ent)}_s$ in its dependence on the choice of the initial coin state. We find that $P^{(ent)}_s$ is bounded in the same way as $P_s^{(sep)}$ displayed in Figure~\ref{fig1:2}. The maximum is obtained for states satisfying the condition $|h_{(++)}|^2 + |h_{(--)}|^2 = 1$, while the minimum is reached when $h_{(++)}=h_{(--)}=0$.}
\label{fig2:2}
\end{center}
\end{figure}

The dependence of the probability $P^{(ent)}_s$ on the initial coin state is illustrated in Figure~\ref{fig2:2}. We find that the probability to be on the same side for entangled initial coin states $P^{(ent)}_s$ satisfies exactly the same bounds as the probability $P^{(sep)}_s$ derived in the previous Section for separable initial coin states. The maximum value of 3/4 is reached when $|h_{(++)}|^2 + |h_{(--)}|^2 = 1$. In such a case, the initial coin state $|\Psi_C\rangle$ is an eigenstate of the two-particle coin $C_H\otimes C_H$ corresponding to the eigenvalue $+1$. On the other hand, the minimum value 1/4 of the probability $P_s^{(ent)}$ is attained when both $h_{(++)}$ and $h_{(--)}$ vanishes. This corresponds to $|\Psi_C\rangle$ being the eigenstate of the coin $C_H\otimes C_H$ with the eigenvalue $-1$.

Finally, we note that for separable coin states the formula (\ref{Ps:ent}) reduces to Eq. (\ref{Ps:sep:had}) which we have derived in the previous Section. Indeed, for separable states we have the relation
$$
h_{(++)} = h_1^+ h_2^+,\quad h_{(--)} = h_1^- h_2^-,
$$
which together with the normalization (\ref{norm:fact}) implies
\begin{eqnarray}
\nonumber P_s & = & \frac{1}{4}\left(\frac{}{} 1 + 2|h_1^+|^2|h_2^+|^2 + 2|h_1^-|^2|h_2^-|^2 \right) \\
\nonumber & = & \frac{1}{4}\left(\frac{}{} 1 + 2|h_1^+|^2|h_2^+|^2 + 2(1-|h_1^+|^2)(1-|h_2^+|^2) \right)\\
\nonumber & = & \frac{1}{4}\left(2+(2\left|h_1^+\right|^2-1) (2\left|h_2^+\right|^2-1)\right) = P_s^{(sep)}\,.
\end{eqnarray}


\section{Indistinguishable particles}
\label{sec:5}

Let us now briefly discuss the probability to be on the same side $P_s$ for indistinguishable particles. We show that for a particular choice of the initial state of the two bosons or fermions the problem reduces to the case of distinguishable particles with maximally entangled coins.

As the initial state of the quantum walk we choose
$$
|\Psi(0)\rangle = |1_{(0,L)}1_{(0,R)}\rangle\,,
$$
i.e. both particles are initially at the origin with the opposite coin states. Recalling the amplitudes $\psi_i^{(L)}$ $(\psi_i^{(R)})$ for the single particle performing the quantum walk with the initial coin state $|L\rangle$ $(|R\rangle$) we express the state of two bosons and fermions in the following form
\begin{eqnarray}
\label{psi:B:F}
\nonumber |\Psi^{(B)}(t)\rangle & = & \sum_{m,n}\sum_{i,j=L,R}\psi_i^{(L)}(m,t)\psi_j^{(R)}(n,t){\hat{a}_{(m,i)}}^\dagger
{\hat{a}_{(n,j)}}^\dagger|vac\rangle\,,\\
|\Psi^{(F)}(t)\rangle & = & \sum_{m,n}\sum_{i,j=L,R}\psi_i^{(L)}(m,t)\psi_j^{(R)}(n,t)\hat{b}_{(m,i)}^\dagger
\hat{b}_{(n,j)}^\dagger|vac\rangle\,,
\end{eqnarray}
where $|vac\rangle$ denotes the vacuum state. Note that in (\ref{psi:B:F}) both summation indexes $m$ and $n$ run over all possible sites. Using the commutation (\ref{commut}) and anti-commutation (\ref{anticommut}) relations we can restrict the sums in (\ref{psi:B:F}) over an ordered pair $(m,n)$ with $m\geq n$. The resulting wave-function will be symmetric or antisymmetric.

We turn to the joint probabilities $p(m,n,t)$ that after $t$ steps we detect a particle at site $m$ and simultaneously a particle at site $n$, with $m\geq n$. First, for $m\neq n$ we find
\begin{eqnarray}
\nonumber p^{(B,F)}(m,n,t) & = & \left|\langle 1_{(m,i)}1_{(n,j)}|\Psi^{(B,F)}(t)\rangle\right|^2 \\
\nonumber & = & \sum_{i,j = L,R}\left|\psi_i^{(L)}(m,t)\psi_j^{(R)}(n,t)\pm\psi_i^{(R)}(m,t)\psi_j^{(L)}(n,t)\right|^2\,,
\end{eqnarray}
where the $+$ sign on the right hand side corresponds to the bosonic $(B)$ , and the $-$ sign to the fermionic $(F)$.
Comparing these expressions with the results for Bell states (\ref{Pent}) we identify the relation
\begin{equation}
\label{P:B:F:psi}
p^{(B)}(m,n,t) = 2 p^{(\psi^+)}(m,n,t)\,,\quad  p^{(F)}(m,n,t) = 2 p^{(\psi^-)}(m,n,t)\,.
\end{equation}
For $m=n$ we obtain for bosons
\begin{eqnarray}
\nonumber p^{(B)}(m,m,t) & = & \left|\langle 2_{(m,L)}|\Psi^{(B)}(t)\rangle\right|^2+\left|\langle 2_{(m,R)}|\Psi^{(B)}(t)\rangle\right|^2 + \\
\nonumber & & + \left|\langle 1_{(m,L)}1_{(m,R)}|\Psi^{(B)}(t)\rangle\right|^2\\
\nonumber & = & 2\left|\psi_L^{(L)}(m,t)\psi_L^{(R)}(m,t)\right|^2 + 2\left|\psi_R^{(L)}(m,t)\psi_R^{(R)}(m,t)\right|^2 +\\
\nonumber & & + \left|\psi_L^{(L)}(m,t)\psi_R^{(R)}(m,t) + \psi_R^{(L)}(m,t)\psi_L^{(R)}(m,t)\right|^2\,,
\end{eqnarray}
and for fermions
$$
p^{(F)}(m,m,t) = \left|\langle 1_{(m,L)}1_{(m,R)}|\Psi^{(F)}(t)\rangle\right|^2 = \left|\psi_L^{(L)}(m,t)\psi_R^{(R)}(m,t) - \psi_R^{(L)}(m,t)\psi_L^{(R)}(m,t)\right|^2\,.
$$
We note that relations similar to (\ref{P:B:F:psi}) hold as well for $m=n$. Indeed, we find the following for bosons
\begin{eqnarray}
\label{P:B:psi}
\nonumber p^{(B)}(m,m,t) & = & \frac{1}{2}\sum_{i,j=L,R}\left|\psi_i^{(L)}(m,t)\psi_j^{(R)}(m,t) + \psi_i^{(R)}(m,t)\psi_j^{(L)}(m,t)\right|^2\\
 & = & p^{(\psi^+)}(m,m,t)\,,
\end{eqnarray}
and for fermions
\begin{eqnarray}
\label{P:F:psi}
\nonumber p^{(F)}(m,m,t) & = & \frac{1}{2}\sum_{i,j=L,R}\left|\psi_i^{(L)}(m,t)\psi_j^{(R)}(m,t) - \psi_i^{(R)}(m,t)\psi_j^{(L)}(m,t)\right|^2\\
 & = & p^{(\psi^-)}(m,m,t)\,.
\end{eqnarray}

Finally, we derive the probability $P_s$ that the bosons (or fermions) are on the same side of the line. As we have already discussed, for indistinguishable particles we have used the formula (\ref{P:same:indist}) where the summation is restricted to an ordered pair $(m,n)$ with $m\geq n$. However, using the results of (\ref{P:B:F:psi}), (\ref{P:B:psi}) and (\ref{P:F:psi}) we can replace $p^{(B,F)}(m,n,t)$ by $p^{(\psi^\pm)}(m,n,t)$ in (\ref{P:same:indist}) and extend the summation over all pairs of $m$ and $n$. Hence, we find that
$$
P^{(B)}_s(t) = P^{(\psi^+)}_s(t)\,,\qquad P^{(F)}_s(t) = P^{(\psi^-)}_s(t)\,.
$$
In summary, the results for bosons (resp. fermions) are the same as for distinguishable particle which have started the quantum walk with entangled coin state $|\psi^+\rangle$ (resp. $|\psi^-\rangle$). This is a direct consequence, of course, of the required symmetry properties of two-particle
boson and fermion states. We note that also the fact that the particles have started the walk from the same lattice point is important. However, when the two indistinguishable particles start the walk spatially separated their evolution differs from that of distinguishable particles with entangled coin states \cite{stef:meeting}. Indeed, indistinguishability starts to play a role when the wave-functions begin to overlap, whereas entanglement is a non-local property.


\section{Quantum walks with $\delta$-interactions}
\label{sec:6}

We have seen in the preceding sections that entanglement in two-particle non-interacting quantum walks cannot break the limit of probabilities we found for separable particles. A natural question arises: What happens if we consider interacting particles? This motivates us to introduce the concept of two-particle quantum walks with $\delta$-interaction. To do that, we change the factorized time evolution operator defined in (\ref{qw:2part:factor}). In the original time evolution the coin was the same factorized coin in all lattice point pairs $(m, n)$, in the $\delta$-interaction quantum walk we change the coin to a non-factorized one $C_{\delta}$, when the particles are at the same lattice point $m=n$.

\begin{figure}
\begin{center}
\includegraphics[width=0.7\textwidth]{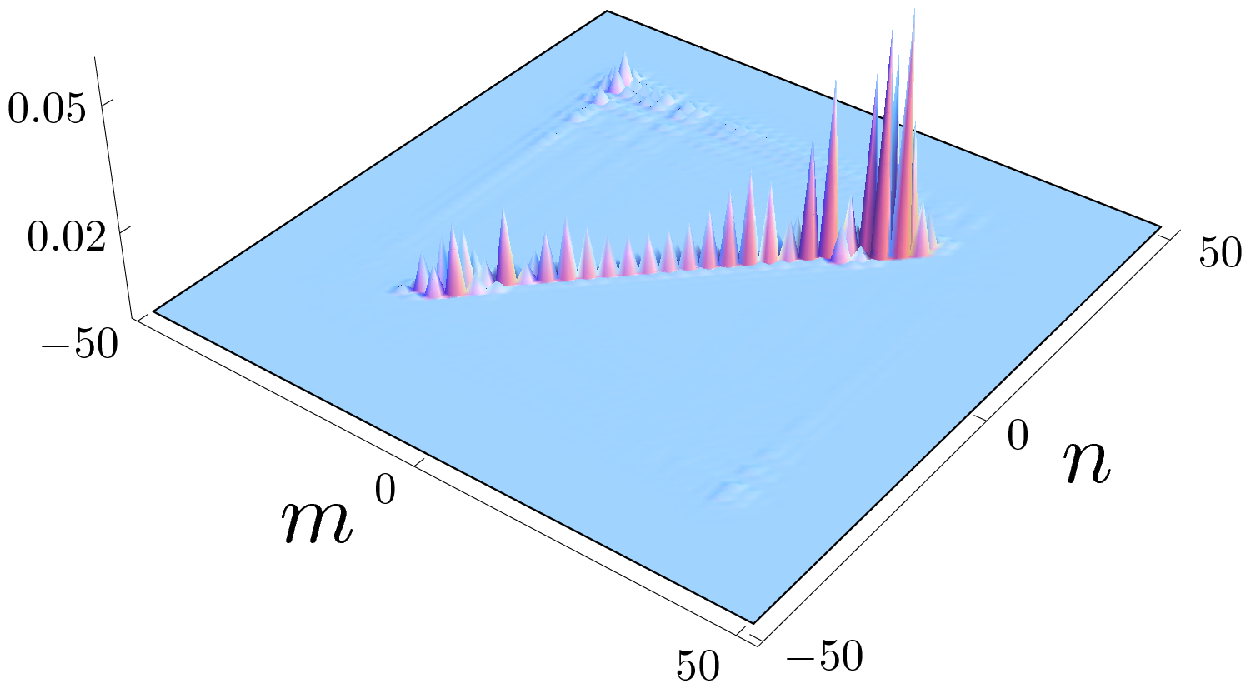}\vspace{12pt}
\includegraphics[width=0.65\textwidth]{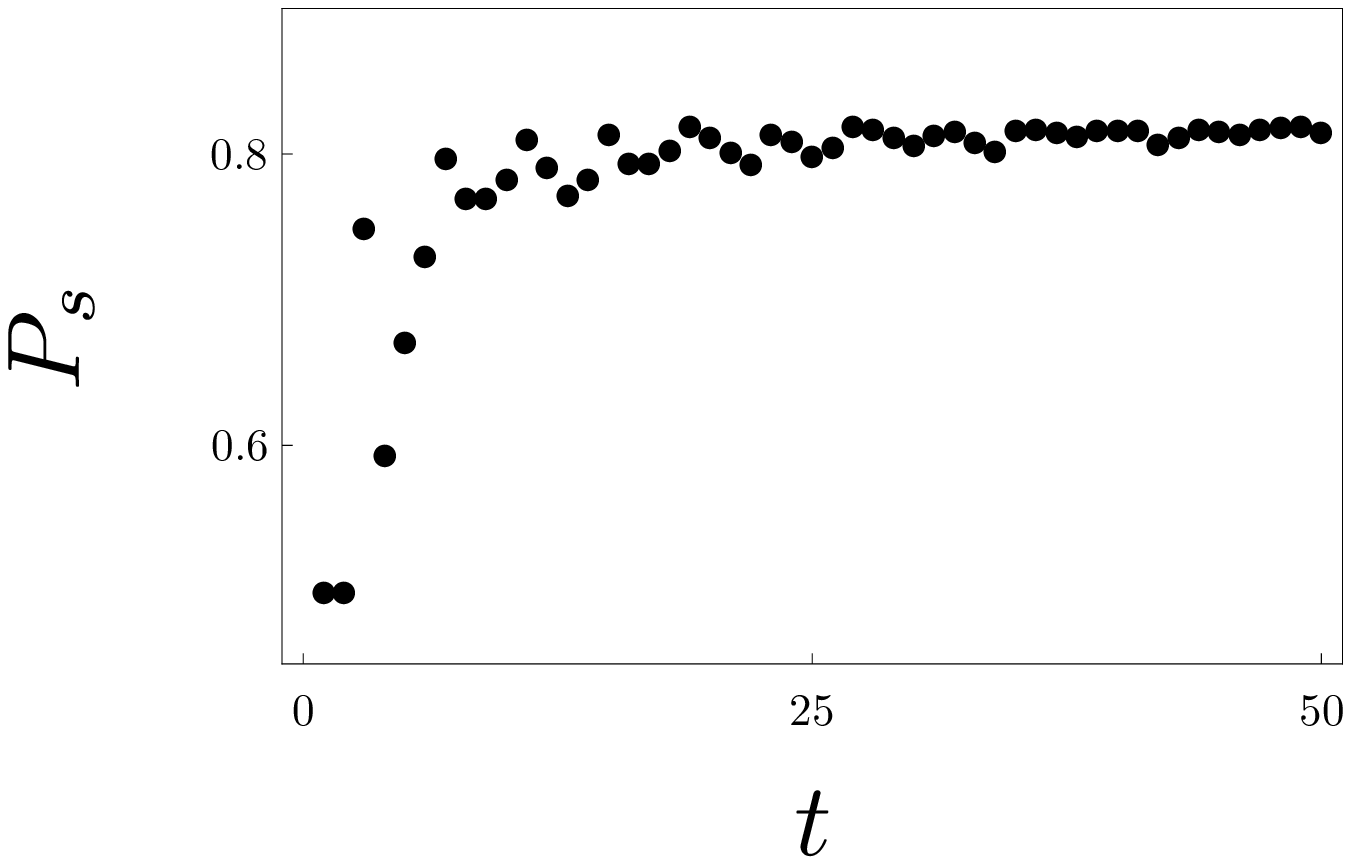}
\caption{Joint probability distribution (upper plot) and the probability to be on the same side of the lattice $P_s$ (lower plot) for two interacting particles performing a quantum walk on a line. The $\delta$-interaction coin $C_{\delta}$ is realized by a matrix (\ref{coin_delta}). As the initial coin state we have chosen one of the Bell states, namely $|\Psi_C\rangle = |\phi^-\rangle$. The resulting joint probability distribution is mostly concentrated on the diagonal, as can be seen from the upper plot. Consequently, the particles are very likely to be on the same side of the lattice. Indeed, the lower plot indicates that the asymptotic value of the probability $P_s$ exceeds 0.8.}
\label{fig6}
\end{center}
\end{figure}

Considering the above, we define the unitary time evolution operator for quantum walks with $\delta$-interacting particles on a line as
$$
 U_{\delta}  =  S_{1 2}(\bar{P_{\delta}} \otimes (C_H  \otimes C_H)) + S_{1  2}(P_{\delta} \otimes C_{\delta})\,,
$$
where $P_{\delta}$ is the projector on the joint position state
$$
 P_{\delta}  =  \sum_m | m \rangle | m \rangle \langle m | \langle m |  \,,
$$
and
$$
 \bar{P_{\delta}}  =  I_{P_{12}} - P_{\delta} \,.
$$
As an example, we consider the entangling $\delta$-interaction coin $C_{\delta}$ of the following form
\begin{equation}
C_{\delta} = \frac{1}{2}\left(
\begin{array}{cccc}
 1 & 1 & 1 & 1 \\
 1 & -1 & -1 & 1 \\
 -1 & 1 & -1 & 1 \\
 -1 & -1 & 1 & 1
\end{array}
\right).
\label{coin_delta}
\end{equation}
In Figure~\ref{fig6} we present the results of a numerical simulation of the corresponding quantum walk with $\delta$-interaction. The initial coin state was chosen to be the Bell state $|\phi^-\rangle$. From the upper plot we find that the joint probability distribution is concentrated on the diagonal, thus the particles are likely to be found on the same side. The lower plot clearly indicate that quantum walks on a line with $\delta$-interactions can break the upper limit of $P_s=3/4$ which we have derived for non-interacting particles.


\section{Conclusions}
\label{sec:7}

We have analyzed the two-particle quantum walk on a line focusing on the directional correlations between the particles. The directional correlation of two non-interacting particles on the line is shown to be confined in an interval, independent of wether the initial state is entangled or not. The bounds of the interval are reached when the initial states coincide with the eigenstates of the coin operator.

Introducing a $\delta$-interaction one can exceed the limit we derived for non-interacting particles. The $\delta$-interaction breaks the translational symmetry, thus new analytical tools are needed to investigate the properties of the introduced model. In the picture of the joint time evolution, this scheme could be considered as an inhomogeneous two-dimensional quantum walk, where the coin is changed on the diagonal line $m=n$.


\ack

The financial support by MSM 6840770039, M\v SMT LC 06002 and the Czech-Hungarian cooperation project (KONTAKT,CZ-11/2009) is gratefully acknowledged. SMB thanks the Royal Society and the Wolfson Foundation for their support.


\appendix

\section{Asymptotic probability distribution for a quantum walk with two entangled particles}
\label{appendix}

In this appendix we derive the asymptotic probability density for a quantum walk on a line with two particles for an arbitrary initial coin state $|\Psi_C\rangle$. We make use of the close relation between the two-particle walk on a line and a single-particle walk on a plane discussed in Section~\ref{sec:2}. We then employ the weak limit theorem \cite{Grimmett}.

The time-evolution of the Hadamard walk on a plane is in the Fourier representation determined by the propagator
$$
\widetilde{U}_{12}(k_1,k_2) = \widetilde{U}_1(k_1)\otimes \widetilde{U}_2(k_2)\,.
$$
Here, $\widetilde{U}_j(k)$ denotes the single-particle propagator of the Hadamard walk on a line, which is given by
$$
\widetilde{U}_j(k) = \mathcal{D}\left(e^{-i k},e^{i k}\right)\cdot C_H\,.
$$
Since $\widetilde{U}_{12}(k_1,k_2)$ has a structure of a tensor product of two unitary matrices we write its eigenvalues in the form
\begin{equation}
\label{app:lambda}
\lambda_{ij}(k_1,k_2) = e^{i\omega_{ij}(k_1,k_2)} = e^{i(\omega_i(k_1)+\omega_j(k_2))},\ i,j=1,2\,,
\end{equation}
where $e^{i\omega_i(k)}$ are the eigenvalues of the matrix $\widetilde{U}_j(k)$. Their phases $\omega_i(k)$ are determined by
\begin{equation}
\label{app:omega}
\omega_1(k) = \arcsin\left(\frac{\sin{k}}{\sqrt{2}}\right),\quad \omega_2(k) = \pi-\omega_1(k)\,.
\end{equation}
Similarly, we write the corresponding eigenvectors of $\widetilde{U}_{12}(k_1,k_2)$ in the form of a tensor product
$$
v_{ij}(k_1,k_2) = v_i(k_1)\otimes v_j(k_2)
$$
of the eigenvectors of the matrices $\widetilde{U}_j(k_j)$
\begin{eqnarray}
\label{app:eigvec}
\nonumber v_1(k) & = & \frac{1}{\sqrt{n_1(k)}}\left(e^{i k},\sqrt{2}e^{i\omega_1(k)}-e^{ik}\right)^T\,,\\
v_2(k) & = & \frac{1}{\sqrt{n_2(k)}}\left(-e^{i k},\sqrt{2}e^{-i\omega_1(k)}+e^{ik}\right)^T\,.
\end{eqnarray}
The normalization of the eigenvectors is given by
\begin{eqnarray}
\nonumber n_1(k) & = & 2\left(1+\cos^2{k} - \cos{k}\sqrt{1+\cos^2{k}}\right)\,,\\
\nonumber n_2(k) & = & 2\left(1+\cos^2{k} + \cos{k}\sqrt{1+\cos^2{k}}\right)\,.
\end{eqnarray}

The weak limit theorem \cite{Grimmett} states that the cumulative distribution function equals
\begin{equation}
\label{app:cdf}
F\left(\widetilde{x}_1,\widetilde{x}_2\right) = \sum_{i,j=1}^2 \int_{\nabla\omega_{i,j}^{-1}\left((-\infty,\widetilde{x}_1)\times(-\infty,\widetilde{x}_2)\frac{}{}\right)} d\mu_{ij}\,,
\end{equation}
where we have denoted $\widetilde{x}_i = \frac{x_i}{t}$. The probability measure $\mu_{ij}$ is determined by
$$
\mu_{ij} = \left|\left(v_{ij}(k_1,k_2),\psi_C\right)\right|^2 \frac{dk_1}{2\pi}\frac{dk_2}{2\pi}\,.
$$
The four-component vector $\psi_C$ corresponds to the initial state of the coin $|\Psi_C\rangle$. From the explicit form of the eigenvectors $v_{ij}(k_1,k_2)$ given in (\ref{app:eigvec}) we find that the probability measures $\mu_{ij}$ equal
\begin{eqnarray}
\label{app:measures}
\nonumber \mu_{ij} & = & \frac{1}{4}\left[\frac{}{} 1 + (-1)^{i+1}\left(\frac{}{} C_1 {\cal C}(k_1) + S_1 {\cal S}(k_1)\right) + \right.\\
\nonumber & & + (-1)^{j+1}\left(\frac{}{} C_2 {\cal C}(k_2) + S_2 {\cal S}(k_2)\right) + \\
\nonumber & & + (-1)^{i+j}\left(\frac{}{} C_{12} {\cal C}(k_1) {\cal C}(k_2) + S_{12} {\cal S}(k_1){\cal S}(k_2) + \right. \\
& & \left.\left.+ X_1 {\cal C}(k_1) {\cal S}(k_2) + X_2 {\cal S}(k_1) {\cal C}(k_2)\frac{}{}\right)\right]\frac{dk_1}{2\pi}\frac{dk_2}{2\pi}\,.
\end{eqnarray}
Here, we have used the notation
$$
{\cal C}(k) = \frac{\cos{k}}{\sqrt{1+\cos^2{k}}},\quad {\cal S}(k) = \frac{\sin{k}}{\sqrt{1+\cos^2{k}}}\,,
$$
to shorten the formulas. The coefficients $C,\ S$ and $X$ entering the expressions (\ref{app:measures}) can be determined from the initial state of the coin $|\Psi_C\rangle$.

To obtain the cumulative distribution function (\ref{app:cdf}) we also have to find the integration domains. These are determined by the gradients of the phases $\omega_{i,j}(k_1,k_2)$ of the eigenvalues of the propagator $\widetilde{U}_{12}(k_1,k_2)$. From their explicit form given in (\ref{app:lambda}) and (\ref{app:omega}) we find that the gradients are
$$
\nabla\omega_{ij}(k_1,k_2) = \left((-1)^{i+1}{\cal C}(k_1),\ (-1)^{j+1} {\cal C}(k_2)\right)\,.
$$
Using the above derived results and the substitution
$$
{\cal C}(k_i) = \frac{\cos{k_i}}{\sqrt{1+\cos^2{k_i}}} = q_i,\quad dk_i = \frac{dq_i}{(1-q_i^2)\sqrt{1-2q_i^2}}\,,
$$
we can simplify the cumulative distribution function into the form
$$
F(\widetilde{x}_1,\widetilde{x}_2) = \frac{1}{\pi^2}\int\limits_{-\frac{1}{\sqrt{2}}}^{\widetilde{x}_1}\frac{dq_1}{(1-q_1^2)\sqrt{1-2q_1^2}}\int\limits_{-\frac{1}{\sqrt{2}}}^{\widetilde{x}_2}\frac{dq_2}{(1-q_2^2)\sqrt{1-2q_2^2}}
\left[\frac{}{}1 -C_1 q_1 - C_2 q_2 + C_{12}q_1 q_2\right]\,.
$$
With the help of the relation
$$
p(x,y) = \frac{\partial^2 F}{\partial x\partial y}
$$
between the cumulative distribution $F(x,y)$ and the probability density $p(x,y)$ we find that the later one is given by
$$
p(x_1,x_2,t) = \frac{1}{\pi^2(1-\frac{x_1^2}{t^2})\sqrt{1-2\frac{x_1^2}{t^2}}(1-\frac{x_2^2}{t^2})\sqrt{1-2\frac{x_2^2}{t^2}}}\left[\frac{}{}1 - C_1 \frac{x_1}{t} - C_2 \frac{x_2}{t} + C_{12}\frac{x_1x_2}{t^2}\right]\,.
$$
Finally, we give the explicit form of the coefficients $C_1,\ C_2$ and $C_{12}$. We find that they have a particularly simple form in the basis formed by the tensor product of eigenvectors of the Hadamard coin $|\chi^\pm\rangle$, which have been given in (\ref{basis:had}). With the decomposition of the initial coin state in the Hadamard basis as given in (\ref{init:2part:had})
we obtain the following expressions for the coefficients $C_{1,2}$ and $C_{12}$:
\begin{eqnarray}
\nonumber C_1 & = & \sqrt{2}\left(|h_{(++)}|^2+|h_{(+-)}|^2-|h_{(-+)}|^2-|h_{(--)}|^2\right)\,,\\
\nonumber C_2 & = & \sqrt{2}\left(|h_{(++)}|^2+|h_{(-+)}|^2-|h_{(+-)}|^2-|h_{(--)}|^2\right)\,,\\
\nonumber C_{12} & = & 2\left(|h_{(++)}|^2+|h_{(--)}|^2-|h_{(+-)}|^2-|h_{(-+)}|^2\right)\,.
\end{eqnarray}


\bibliography{bibliography}{}
\bibliographystyle{iopart-num}

\end{document}